\input{aipcheck}

\documentclass[
    ,final            % use final for the camera ready runs
  ]
  {aipproc}

\layoutstyle{6x9}

\newcommand {\pom} {I\!\! P}
\begin{document}
\input{psfig.sty}
\title{Multigap Diffraction at LHC}


\author{Konstantin Goulianos}{
  address={The Rockefeller University, New York, NY 10021, USA}
}

\begin{abstract}
The large rapidity interval available at the Large Hadron Collider (LHC) 
offers an arena in which the QCD aspects of diffraction may be explored 
in an environment free of gap survival 
complications using events with multiple rapidity gaps.
\end{abstract}

\maketitle
\vspace{-17em}
%\hfill{DF/PUB/JET/PUBLIC/7726}

\begin{center}Presented at DIS-2005, XIII$^{th}$ International Workshop on Deep Inelastic Scallering,\\
April 27 - May 1 2005, Madison, WI, U.S.A.
\end{center}
\vspace{11em}

%%%%%%%%%%%%%%%%%%%%%%%%%%%%%%%%%%%%%%%%%%%%
%% MAINMATTER
%%%%%%%%%%%%%%%%%%%%%%%%%%%%%%%%%%%%%%%%%%%%

\section{Soft Diffraction}
Diffractive processes are characterized 
by large rapidity gaps, defined as regions of 
(pseudo)rapidity~\cite{pseudo}
in which there is no particle production. 
Diffractive gaps are presumed to be produced by the exchange of  
a color singlet quark/gluon object 
with vacuum quantum numbers referred to as 
the {\em Pomeron}~\cite{lathuile,ismd04} (the present paper contains 
excerpts from these two references). 

Traditionally diffraction had been treated in Regge theory using 
an amplitude based on a simple Pomeron pole and factorization.
This approach was successful at $\sqrt s$ 
energies below $\sim 50$ GeV~\cite{physrep}, but as collision
energies increased to reach $\sqrt s=$1800 GeV at the Fermilab Tevatron 
the SD cross section was found to be suppressed by a factor of 
$\sim {\cal{O}}(10)$ relative to the Regge-based prediction~\cite{sd}. 
This blatant breakdown of factorization was traced back to 
the energy dependence of the Regge theory  
$\sigma_{sd}^{tot}(s)$,
\begin{equation}
%\hbox{ Regge: }
d\sigma_{sd}(s,M^2)/dM^2\sim s^{2\epsilon}/(M^2)^{1+\epsilon},
\label{eq:reggeM2}
\end{equation} 
which is faster than that of $\sigma^{tot}(s)\sim s^\epsilon$, 
so that at high $\sqrt s$ unitarity would have to be 
violated if factorization held. 

In contrast to the Regge theory prediction of Eq.~(\ref{eq:reggeM2}), the
measured SD $M^2$-distribution shows no explicit
$s$-dependence ($M^2$-scaling) over a region of $s$ spanning six orders of 
magnitude~\cite{GM}. Thus, factorization appears to {\em yield} to 
$M^2$-scaling. This is a property built into the
{\em Renormalization Model} of hadronic diffraction, in which the 
Regge theory Pomeron flux is renormalized to unity~\cite{R}.
\vspace{-2em}
\unitlength 0.95in
\thicklines
\begin{center}
\begin{picture}(6,1)(1,0)
\put(1,0){\line(2,0){6}}
\multiput(2,0)(2,0){3}{\oval(0.8,0.5)[t]}
\put(1.9,-0.25){$\eta_1'$}
\put(2.9,-0.25){$\eta_2$}
\put(3.9,-0.25){$\eta_2'$}
\put(4.9,-0.25){$\eta_3$}
\put(5.9,-0.25){$\eta_3'$}
\put(2.8,0.5){$\Delta \eta_2$}
%\put(3.35,-0.65){$\Delta \eta\equiv \sum_{i=1}^4\Delta y_i$}
\put(4.8,0.5){$\Delta \eta_3$}
\put(1.25,-0.5){$t_1$}
\put(2.9,-0.5){$t_2$}
\put(4.9,-0.5){$t_3$}
\put(6.7,-0.5){$t_4$}
\put(1.15,0.5){$\Delta \eta_1$}
\put(1.85,0.5){$\Delta \eta'_1$}
\put(3.85,0.5){$\Delta \eta'_2$}
\put(5.85,0.5){$\Delta \eta'_3$}
\put(6.6,0.5){$\Delta \eta_4$}
\end{picture}
\end{center}
\vspace{2em}
\begin{figure}[h]
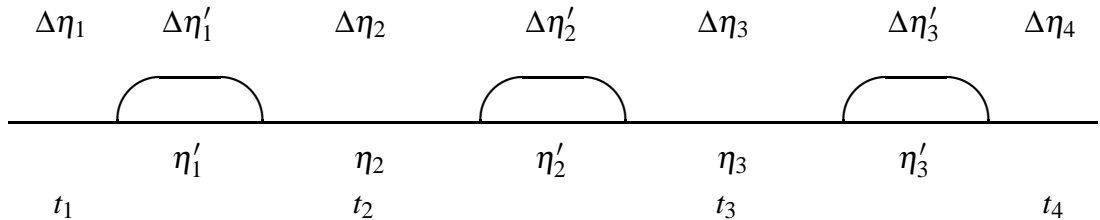

\caption{Average multiplicity $dN/d\eta$ versus $\eta$ for a process with four rapidity gaps $\Delta \eta_{i=1-4}$.} 
\label{fig:soft}
\end{figure}

In a QCD inspired approach, the renormalization model was extended
to central and multigap diffractive processes~\cite{corfu}, an example of  
which is the four-gap process shown schematically 
in Fig~\ref{fig:soft}.
In this approach cross sections depend on the 
number of wee partons~\cite{Levin} and therefore the $pp$ total cross section 
is given by 
\begin{equation}
\sigma_{pp}^{tot}=\sigma_0\cdot e^{\epsilon\Delta\eta'},
\label{totDeta}
\end{equation}
where $\Delta \eta'$ is the rapidity region in which there is particle production. Since, from the optical theorem, $\sigma^{tot}\sim {\rm Im\,f^{el}}(t=0)$, 
the full parton model amplitude may be written as 
\begin{equation}
{\rm Im\,f^{el}}(t,\Delta\eta)\sim e^{({\epsilon}+\alpha't)\Delta \eta},
\label{eq:fPM}
\end{equation}
\noindent where $\alpha't$ is 
a simple parameterization of the $t$-dependence of the amplitude. 
On the basis of this amplitude, the
cross section of the four-gap process of Fig.~\ref{fig:soft} takes the form 

\begin{equation}\
\frac{d^{10}\sigma^D}{\Pi_{i=1}^{10}dV_i}=
N^{-1}_{gap}\; 
\underbrace{
F^2_p(t_1)F^2_p(t_4)
\Pi_{i=1}^4\left\{e^{[\epsilon+\alpha't_i]\Delta\eta_i}\right\}^2
}_{\hbox{gap probability}}\;
\times\kappa^4\left[\sigma_0\,e^{\epsilon\sum_{i=1}^3\Delta\eta'_i}\right],
\label{eq:diffPM}
\end{equation}
\noindent where the term in square brackets is the $pp$  
total cross section at the reduced $s$-value, defined 
through $\ln (s'/s_0)=\sum_i\Delta \eta_i'$, 
$\kappa$ (one for each gap) is the QCD color factor for gap formation, 
the gap probability is the amplitude squared for elastic scattering 
between two diffractive clusters or between a diffractive cluster and a 
surviving proton with form factor $F^2_p(t)$,
and $N_{gap}$ is the (re)normalization factor defined as 
the gap probability integrated 
over all 10 independent variables $t_i$, $\eta_i$, $\eta'_i$, and 
$\Delta\eta\equiv \sum_{i=1}^4\Delta\eta_i$. 

The renormalization 
factor $N_{gap}$ is a function of $s$ only.
The color factors are $c_g=(N_c^2-1)^{-1}$ 
and $c_q=1/N_c$ for gluon and quark color-singlet exchange, respectively.  
Since the reduced energy cross section is properly normalized, the 
gap probability is (re) normalized to unity. The quark to gluon 
fraction, and thereby the Pomeron intercept parameter $\epsilon$ 
may be obtained from the inclusive parton distribution 
functions (PDFs)~\cite{lathuile}. Thus, normalized differential multigap cross 
sections at $t=0$ may be fully derived from inclusive PDFs and QCD color 
factors without any free parameters.

The exponential dependence of the cross section on $\Delta\eta_i$ leads to 
a renormalization factor $\sim s^{2\epsilon}$ independent of the number of 
gaps in the process. This remarkable property of the renormalization model, 
which was confirmed in two-gap to one-gap cross section ratios measured by 
the CDF Collaboration (see~\cite{lathuile}), suggests that multigap
diffraction can be used as a tool for exploring the QCD aspects of 
diffraction in an environment free of rapidity gap suppression effects.
The LHC with its large rapidity coverage provides the ideal arena 
for such studies.  

\section{Hard Diffraction}
%\subsection{QCD factorization in diffraction}
Hard diffraction processes are those in which there is a hard 
partonic scattering in addition to the diffractive rapidity gap. 
SD/ND ratios for $W$, dijet, $b$-quark, and $J/\psi$ production 
at $\sqrt s=$1800 GeV measured by the CDF Collaboration are 
approximately equal ($\sim 1\%$), indicating that the rapidity gap 
formation probability is largely {\em flavor independent}.
However, the SD structure function measured from dijet production 
is suppressed by $\sim{\cal{O}}(10)$ relative to expectations based
on diffractive PDFs measured from 
diffractive DIS at HERA.
 
A modified version of our QCD approach to soft diffraction 
can be used to describe hard diffractive processes and has been  
applied to diffractive DIS at HERA, 
$\gamma^*+p\rightarrow p+Jet+X$, and diffractive dijet production at the 
Tevatron, $\bar p+p\rightarrow\bar p+\hbox{dijet}+X$ in~\cite{ismd04}. 
The hard process generally involves several color ``emissions''
from the surviving proton, the sum of which comprises 
a color singlet exchange with vacuum quantum numbers. 
Two of these emissions are of special interest, 
one at $x=x_{Bj}$ from the proton's hard PDF at scale $Q^2$, 
which causes the hard scattering, and another at $x=\xi$ 
(fractional momentum loss of the diffracted nucleon) from the soft 
PDF at $Q^2\approx 1$ GeV$^2$, which neutralizes the exchanged color 
and forms the rapidity gap. Neglecting the $t$-dependence, 
the diffractive structure function could then be 
expressed as the product of the inclusive 
hard structure function and the soft parton 
density at $x=\xi$,

\begin{equation}F^D(\xi,x,Q^2)=
\frac{A_{\rm norm}}{\xi^{1+\epsilon}}\cdot c_{g,q}\cdot F(x,Q^2)
%=\frac{A_{\rm norm}}{\xi^{1+\epsilon}}c_{g,q}\frac{C(Q^2)}{(\beta\xi)^{\lambda(Q^2)}}
\Rightarrow
\frac{A_{\rm norm}}{\xi^{1+\epsilon+\lambda(Q^2)}}
\cdot c_{g,q}\cdot \frac{C(Q^2)}{\beta^{\lambda(Q^2)}},
\label{eq:F2D3}
\end{equation}
where $c_{g,q}$ are QCD color factors, 
$\lambda$ is the parameter of a power law fit to the hard structure 
function in the region $x<0.1$,
$A_{\rm norm}$ is a normalization factor, and $\beta\equiv x/\xi$.

At high $Q^2$ at HERA, where factorization is expected to 
hold~\cite{R,JCollins}, $A_{\rm norm}$ is the nominal normalization 
factor of the soft PDF. This factor is constant, leading to two 
important predictions, which are confirmed by the data:

i) The Pomeron intercept in diffractive DIS (DDIS) 
is $Q^2$-dependent and equals 
the average value of the soft and hard intercepts:
\begin{equation}
\alpha^{DIS}_{\pom}=1+\lambda(Q^2),\;\;\;\;
\alpha^{DDIS}_{\pom}=1+\frac{1}{2}\left[\epsilon+\lambda(Q^2)\right]\nonumber
\label{eq:heraintercept}
\end{equation}

ii) The ratio of DDIS to DIS structure functions at fixed $\xi$ 
is independent of $x$ and $Q^2$:
\hspace*{-5em}\begin{equation}
R\left[\frac{F^D(\xi,x,Q^2)}{F^{ND}(x,Q^2)}\right]_{\rm HERA}=
\frac{A_{\rm norm}\cdot c_q}{\xi^{1+\epsilon}}=
\frac{\rm const}{\xi^{1+\epsilon}}
\label{eq:Rhera}
\end{equation}     

At the Tevatron, where high soft parton densities lead to saturation,
$A_{\rm norm}$ must be renormalized to
\begin{equation}
A^{\rm Tevatron}_{\rm renorm}=1/\int^{\xi=0.1}_{\xi_{min}}
\frac{d\xi}{\xi^{1+\epsilon+\lambda}}
\propto
\left(\frac{1}{\beta\cdot s}\right)^{\epsilon+\lambda},
\label{eq:tevrenorm}
\end{equation}
where $\xi_{min}=x_{min}/\beta$ and $x_{min}\propto 1/s$.
Thus, the diffractive structure function acquires a term
$\sim (1/\beta)^{\epsilon+\lambda}$, and the ratio of the 
diffractive to inclusive 
structure functions a term $\sim (1/x)^{\epsilon+\lambda}$.
This prediction is confirmed by CDF data,
where the $x$-dependence of the diffractive to inclusive ratio 
was measured to be $\sim 1/x^{0.45}$ (see~\cite{lathuile}). 

A comparison~\footnote{Performed by the author and K. Hatakeyama (see~\cite{lathuile}) using CDF published data and preliminary H1 diffractive parton densities~\cite{heraglue}.}
between the diffractive structure function measured on the proton side 
in events with a leading antiproton to 
expectations from diffractive DIS at HERA showed 
approximate agreement, indicating that factorization is largely 
restored for events that already have a rapidity gap.
Thus, as already mentioned for soft diffraction, 
events triggered on a leading proton at LHC provide an environment 
in which the QCD aspects of diffraction may be explored without 
complications arising from rapidity gap survival.    
\section{Proposed program of multigap diffraction at LHC}
The rapidity span at LHC running at $\sqrt s=14$ TeV is $\Delta\eta=19$ as 
compared to $\Delta\eta=15$ at the Tevatron. This suggests the following 
program for studies of non-suppressed diffraction:
\begin{itemize}
\item Trigger on two forward rapidity gaps  of $\Delta\eta_F\geq 2$ 
(one on each side of the 
interaction point), or equivalently on 
forward protons of fractional longitudinal momentum loss 
$\xi=\Delta p_L/p_L\leq 0.1$, and explore the central rapidity region of 
$|\Delta\eta|\leq 7.5$, which has the same width 
as the entire rapidity region of the Tevatron.
In such an environment, the ratio of the rate of 
dijet events with a gap between jets to that without a gap,  
$gap$+[jet-gap-jet]+$gap$ to $gap$+[jet-jet]+$gap$, should rise 
from its value of $\sim 1$\% at the Tevatron to $\sim 5$\%. 
\item Trigger on one forward gap of $\Delta\eta_F\geq 2$ or on a proton of $\xi<0.1$,
in which case the rapidity gap available for non-suppressed diffractive 
studies rises to 17 units.
\end{itemize}


\begin{thebibliography}{0}
\bibitem{pseudo}
We use {\em pseudorapidity}, $\eta=-\ln\tan\frac{\theta}{2}$, and 
{\em rapidity}, $y=\frac{1}{2}\frac{E+p_L}{E-p_L}$, interchangeably.
%\bibitem{Bj} J.D. Bjorken, {\em Phys. Rev.} {\bf D47}, 101 (1993).
%\bibitem{Collins}P.D.B. Collins, in {\em An Introduction to Regge Theory and High Energy Physics} (Cambridge University  Press 1977).
%\bibitem{lathuile} K. Goulianos, ``Hadronic Diffraction: Where do we Stand?,'' Les Rencontres de Physique de la Vall\'{e} d'Aoste, La Thuile, Aosta Valley, Italy, February 29 - March 6, 2004; e-Print Archive: hep-ph/0407035.
\bibitem{lathuile}K.~Goulianos, ``Hadronic Diffraction: Where do we Stand?,'' in {\em La Thuile 2004, Results and Perspectives in Particle Physics},
edited by M.~Greco, Proc. of 
Les Rencontres de Physique de la Vall\'{e} d'Aoste, La Thuile, Aosta Valley, Italy, February 29 - March 6, 2004, pp. 251-274; e-Print Archive: hep-ph/0407035.
\bibitem{ismd04}K. Goulianos, {\em Nucl. Phys. (Proc. Suppt.) B}, {\bf 146}, 166 (2005).
%\bibitem{dd}T. Affolder {et al.} (CDF Collaboration), {\em Phys. Rev Lett.} {\bf 87}, 141802 (2001). 
%\bibitem{idpe}D. Acosta {\em et al.} (CDF Collaboration), {\em Phys. Rev. Lett.} {\bf 93}, 141601 (2004).  
%\bibitem{sdd}D. Acosta {et al.} (CDF Collaboration), {\em Phys. Rev Lett.} {\bf 91}, 011802 (2003). 
\bibitem{physrep}K. Goulianos, {\em Phys. Reports}, {\bf 101}, 171 (1983).
\bibitem{sd}F. Abe {et al.} (CDF Collaboration), {\em Phys. Rev. D}, {\bf 50}, 5535 (1994).
\bibitem{GM}K. Goulianos and J. Montanha, {\em Phys. Rev. D}, {\bf 59}, 114017 (1999).
\bibitem{R}K. Goulianos,
{\em Phys. Lett. B}, {\bf 358}, 379 (1995); 
Erratum-{\em ib.} {\bf 363}, 268 (1995).
\bibitem{corfu}K. Goulianos, in {\em Diffraction in QCD}, 
Corfu Summer Institute on Elementary Particle Physics, Corfu, 
Greece, 31 Aug - 20 Sep 2001;
e-print Archive: hep-ph/0203141.
\bibitem{Levin}E. Levin, ``An Introduction to Pomerons,'' 
Preprint DESY 98-120.
%\bibitem{CMG}R.J.M. Covolan, J. Montanha, and K. Goulianos, {\em Phys. Lett.} {\bf B389} (1996) 176.
\bibitem{H1_lambda}A. Petrukhin (H1 Collaboration), ``Measurement of the Inclusive DIS Cross Section at Low 
$Q^2$ and High $x$ Using Events with Initial State Radiation,'' 
presented at DIS2004, 14-18 April 2004, Slovakia.
\bibitem{heraglue}F.~P. Schilling (H1 Collaboration), 
``Measurement and NLO DGLAP
  QCD Interpretation of Diffractive Deep-Inelastic Scattering at HERA,''
  submitted to 31$^{st}$ International Conference on High Energy Physics,
  ICHEP02, Amsterdam, The Netherlands, Jul. 24$-$31, 2001 (abstract 980).
\bibitem{JCollins}J. Collins,  
{\em J. Phys. G}, {\bf 28}, 1069 (2002); e-Print Archive: hep-ph/0107252. 
\end{thebibliography}
\end{document}